\documentclass[prb,twocolumn,showpacs,superscriptaddress,preprintnumbers,amssymb]{revtex4-2}
\usepackage{graphicx}
\usepackage{color}
\usepackage{dcolumn}
\usepackage{bm}
\usepackage{graphicx}
\usepackage{mathtools}
\usepackage{physics}
\usepackage{hyperref}

\renewcommand{\v}[1]{{\bf #1}}

\newcommand{\beq}{\begin{equation}}
\newcommand{\eeq}{\end{equation}}
\newcommand{\beqn}{\begin{eqnarray}}
\newcommand{\eeqn}{\end{eqnarray}}

\newcommand{\sgn}{{\rm sgn}}

\newcommand{\ua}{\uparrow}
\newcommand{\da}{\downarrow}
\newcommand{\ra}{\rightarrow}

\newcommand{\cL}{ {\cal L} }
\newcommand{\cP}{ {\cal P} }

\newcommand{\cS}{ {\cal S} }

\newcommand{\vect}[1]{{\bm{#1}}}

\newcommand{\ii}{\mathrm{i}}

\newcommand{\hn}{\hat{n}}
\newcommand{\hO}{\hat{O}}

\newcommand{\hP}{\hat{P}}

\newcommand{\llangle}{\langle\!\langle}
\newcommand{\rrangle}{\rangle\!\rangle}

\newcommand{\SU}{\mathrm{SU}}
\newcommand{\U}{\mathrm{U}}

\newcommand{\calH}{\mathcal{H}}

\newcommand{\calO}{\mathcal{O}}
\newcommand{\calP}{\mathcal{P}}

\DeclareUnicodeCharacter{2215}{\ensuremath{/}}

\newcommand{\kket}[1]{\mbox{$| #1 \rangle\!\rangle$}}
\newcommand{\bbra}[1]{\mbox{$\langle\!\langle #1 |$}}

\newcommand{\cx}[1]{{\color{black} #1}}
\newcommand{\yimu}[1]{\textcolor{black} {#1}}
\definecolor{orange_custom}{rgb}{0.93, 0.47, 0.2}

\begin{document}

\title{Spin Liquid and Superconductivity emerging from Steady States and Measurements }

\author{Kaixiang Su}

\author{Abhijat Sarma}

\affiliation{Department of Physics, University of California,
Santa Barbara, CA 93106, USA}

\author{Marcus Bintz}

\affiliation{Department of Physics, Harvard University,
Cambridge, MA 02138, USA}

\author{Thomas Kiely}

\author{Yimu Bao}

\affiliation{Kavli Institute for Theoretical Physics, University
of California, Santa Barbara, CA 93106, USA}

\author{Matthew P. A. Fisher}

\affiliation{Department of Physics, University of California,
Santa Barbara, CA 93106, USA}

\author{Cenke Xu}

\affiliation{Department of Physics, University of California,
Santa Barbara, CA 93106, USA}

\date{\today}

\begin{abstract}

We demonstrate that, starting with a simple fermion wave function, the steady mixed state of the evolution of a class of Lindbladians, and the ensemble created by strong local measurement of fermion density without post-selection can be mapped to the ``Gutzwiller projected" wave functions in the doubled Hilbert space -- the representation of the density matrix through the Choi-Jamio\l kowski isomorphism. A Gutzwiller projection is a broadly used approach of constructing spin liquid states. For example, if one starts with a gapless free Dirac fermion pure quantum state, the constructed mixed state corresponds to an algebraic spin liquid in the doubled Hilbert space. We also predict that for some initial fermion wave function, the mixed state created following the procedure described above is expected to have a spontaneous ``strong-to-weak" $\U(1)$ symmetry breaking, which corresponds to the emergence of superconductivity in the doubled Hilbert space. We also design the experimental protocol to construct the desired physics of mixed states. 

\end{abstract}

\maketitle

\section{Introduction}

Quantum spin liquids, a class of highly nontrivial disordered phase of quantum spins, 
have been 
the subject of an extremely active subfield of condensed matter physics since their
early theoretical proposal~\cite{anderson1,anderson2}. Despite great progress made in the field, many open questions and challenges remain (for a review, please refer to~\cite{balentsreview1,balentsreview2,zhoureview}). First of all, the theoretical description of spin liquids usually involves a
strongly coupled gauge theory, which is a formidable analytical problem except for certain theoretical limits.
Secondly, though it is certain that spin liquids do 
exist in some elegant theoretical models, e.g. the Kitaev model~\cite{kitaev2006}, numerical simulation of quantum spin liquid on more realistic models often suffers from sign problems as the models that potentially realize the spin liquid usually have geometric frustration. Hence controversies continue to persist about the existence or nature of the spin liquids in various important realistic models. Thirdly, the signal of quantum spin liquid in real correlated materials may be obscured by the inevitable disorders and impurities.

\cx{Rather than the parent Hamiltonian, one can instead focus on the spin liquid wave function. One standard construction of spin liquid wave function is the so-called Gutzwiller projected state~\cite{motrunich2005,ran2007,hermele2008,Grover2010,xuVMC,becca,fisherboseliquid}: \beqn |{\rm SL}\rangle = \prod_{i} \hat{\cP}(\hn_{i,\ua} + \hn_{i,\da} = 1) |f_{i,\ua}, f_{i,\da}\rangle, \eeqn where $|f_{i,\ua}, f_{i,\da}\rangle$ is a simple spin-1/2 fermion state, and it can be often taken as a free fermion state with on average one fermion per-site. The projection $\hat{\cP}$ ensures that there is one and only one fermion per site, which matches the onsite Hilbert space of a spin-1/2 system. } 

\cx{Though the Gutzwiller projection is broadly used as a trial mean field wave functions of spin liquids, it is never exactly realized in condensed matter systems.} In this work we demonstrate that the Gutzwiller wave function can be realized in a completely different set-up: it can be constructed as the steady mixed state of a Lindbladian evolution, or as the ensemble created by strong measurements of local operators, starting with a simple fermion wave function. The mixed state density matrix obtained from both constructions in the doubled Hilbert space (the Choi-Jamio\l kowski representation~\cite{choi,choi2}), becomes exactly a Gutzwiller wave function. 
Predictions of the constructed mixed state will be made based on our theoretical understanding of quantum spin liquids.

We also demonstrate that, sometimes the constructed mixed state density matrix is expected to become a supercondutor in the doubled Hilbert space, which corresponds to the ``strong-to-weak" spontaneous symmetry breaking, a subject that has attracted great interests very recently~\cite{wfdecohere,biswssb,biswssb2,Ogunnaike_2023,huangswssb,yizhiswssb}. These theoretical predictions and expectations can be tested using experimental protocols designed in the supplementary material (SM).

\section{Basic Formalism}

\subsection{Construction with Lindbladian}

We consider the nonunitary Lindbladian evolution of a density matrix. Let us assume that the density matrix at $t = 0$ corresponds to a pure quantum state $\rho_0 = |\Psi_0 \rangle \langle \Psi_0|$ of fermions. For simplicity we will start with a non-interacting spinless fermion wave function $|\Psi_0\rangle$, and we assume that $|\Psi_0\rangle$ is the ground state of a Hamiltonian $H_0$ on a $d-$dim lattice: $ H_0 = \sum_{\langle i,j\rangle} - t_{ij} c^\dagger_i c_j + \mathrm{h.c.} $ We will first assume that $H_0$ is a {\it gapless} free fermion tight-binding model, later we will discuss interactions in $H_0$.

We consider the situation where the jump operators are the local fermion density operators $L_i = \hn_{i} = c^\dagger_i c_i$. The nonunitary evolution of a density matrix under a general Lindbladian reads (we assume that the unitary part of the evolution is absent) \beqn \partial_t \rho(t) = \sum_i L_i \rho L_i^\dagger - \frac{1}{2} \left( L_i^\dagger L_i \rho + \rho L_i^\dagger L_i \right). \eeqn

Exploring physics of mixed-states using the Choi-Jamio\l kowski isomorphism of the density matrix, i.e. representing the density matrix as a pure quantum state in the doubled Hilbert space~\cite{choi,choi2}, has attracted a great deal of interests~\cite{bao2023mixedstatetopologicalordererrorfield, lee2024symmetryprotectedtopologicalphases,wang2023topologicallyorderedsteadystates,dai2023steadystatetopologicalorder,PhysRevB.109.245109,xu2024averageexactmixedanomaliescompatible,arovas,Lin_spinliquid,Lin_spinliquid2}. The Choi-Jamio\l kowski isomorphism maps a density matrix $\rho = \sum_{n} p_n |\psi_n\rangle \langle \psi_n | $ to a state $|\rho\rrangle \sim \sum_n p_n |\psi_{n,L}\rangle |\psi_{n,R}\rangle$. In our current case it is also convenient to take the Choi-Jamio\l kowski representation of the density matrix. The evolution of the Choi state in the doubled Hilbert space is given by \beqn |\rho_t \rrangle \sim e^{t \cL } |\rho_0\rrangle, \eeqn
where \beqn \cL &=& \sum_i \hn_{i,L} \hn_{i,R} - \frac{1}{2}(\hn_{i,L}^2 + \hn_{i,R}^2) \cr\cr &=& \sum_i - \frac{1}{2} (\hn_{i,L} - \hn_{i,R})^2. \eeqn Here $L$ and $R$ label two copies of fermionic modes in the doubled Fock space of the Choi-Jamio\l kowski isomorphism of the density matrix. The initial Choi state $\kket{\rho_0}$ for the pure state $\rho_0 = \ketbra{\Psi_0}$ is the free fermion state with the parent Hamiltonian
\begin{align}
\calH_0 &= H_0(c_{i,L}) + H_0^*(c_{i,R}) \nonumber \\
&= \sum_{\langle i,j\rangle} - t_{ij} c_{i,L}^\dagger c_{j,L} - t^{*}_{ij} c_{i,R}^\dagger c_{j,R} + \mathrm{h.c.}. \label{freeH}
\end{align}
In the limit $t \to \infty$, the steady state $\kket{\rho_\infty}$ satisfies the constraint $\hn_{i,L} - \hn_{i,R} = 0$.


To more explicitly reveal the implication of the constraint $\hn_{i,L} - \hn_{i,R} = 0$,  let us perform a particle-hole (PH) transformation on $c_{i,R}$, and formally relabel the fermions in the $L$, $R$ spaces as ``$\ua$" and ``$\da$": \beqn && c_{i, L} \ra f_{i, \ua}, \ \ c_{i, R}\ra \eta_i f^\dagger_{i,\da}, \cr\cr && \hn_{i,L} \ra \hn_{i,\ua}, \ \ \hn_{i,R} \ra 1 - \hn_{i,\da}, \eeqn where $\eta_i = \pm 1$ can be chosen to depend on the sites $i$. Then in the limit $t \ra \infty$, the constraint $\hn_{i,L} - \hn_{i,R} = 0$ becomes \beqn \hn_{i,\ua} + \hn_{i,\da} = 1, \eeqn which is precisely the {\it Gutzwiller projection}.
%
%
In this case, the steady Choi state is mapped to a spin wave function in the standard spin liquid literature: \beqn |\rho_{\infty}\rrangle = \prod_i \hat{\calP}(\hn_{i,\ua} + \hn_{i,\da} = 1) |\rho_0\rrangle, \label{projwf1} \eeqn where $\hat{\calP}(\hn_{i,\ua} + \hn_{i,\da} = 1)$ is the projector onto the subspace with single occupation.
Since now on every site there is one and only one fermion $f_{i,\alpha}$, the projected wave function $|\rho_\infty\rrangle$ lives in an effective spin-1/2 Hilbert space, with spin-1/2 operators $\hat{S}^\mu_i = \frac{1}{2} f^\dagger_{i,\alpha} \sigma^\mu_{\alpha\beta} f_{i,\beta}$. Under certain conditions, for example, (1) the fermions in $|\Psi_0\rangle$ are at half-filling, i.e. there are on average 1/2 fermions per site; and (2) under particle-hole transformation $H_0^\ast$ becomes $H_0$ with a certain choice of $\eta_i$, then the parent tight-binding Hamiltonian of $|\rho_0\rrangle$ in the doubled Hilbert space enjoys a full effective $\SU(2)$ ``spin" symmetry. For example, if $H_0$ is a tight-binding model on a bipartite lattice with nearest neighbor hopping, then choosing $\eta_i = -1$ on one sublattice will result in a $\SU(2)$ symmetric $\calH_0$ in Eq.~\ref{freeH}. 

\subsection{Construction with measurements}

When the jump operators are Hermitian and all commute with each other, the steady state density matrix also describes the ensemble of wave functions generated by strong measurements on the jump operators {\it without post-selection}. For example, when we perform strong measurement of local densities $\hn_i$ on a pure fermion quantum state $|\Psi_0\rangle$, {\it without} post-selecting the measurement outcomes, the density matrix of the ensemble becomes \beqn \rho = \sum_{\mathbf{n}}  \hat{P}_{\mathbf{n}} \rho_0 \hat{P}_{\mathbf{n}}, \eeqn where $\mathbf{n}$ represents a configuration of fermion numbers $\hn_i$, and $\hat{P}_{\mathbf{n}}$ is a projection operator to the particular configuration. Such an ensemble can be obtained in fermi gas microscope with high spatial resolution~\cite{coldatom3,gross2021quantum,cheuk2015quantum,haller2015single,parsons2015site,edge2015imaging,omran2015microscopic}. \cx{The SM includes a detailed experimental protocol for verifying the predictions made in this work based on $\rho$ generated by measurement. } 

We also note that, construction of spin liquids as the doubled state representation of a mixed state was discussed before~\cite{arovas,Lin_spinliquid,Lin_spinliquid2}, but to the best of our knowledge, the general connection to the Gutzwiller wave function, which is a broadly used construction of spin liquids, has not been explored. 

\begin{figure}[h!]
     \includegraphics[width=0.85\linewidth]{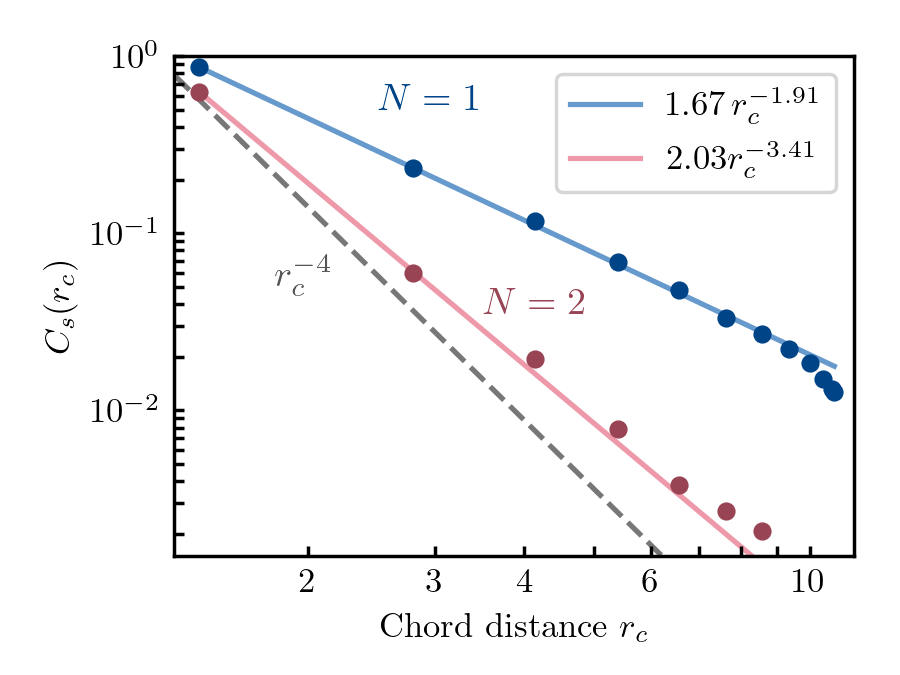}
     \caption{The staggered spin-spin correlation of the projected $\pi$-flux state on the square lattice ($24\times24$ torus), which is a Dirac spin liquid constructed with $N = 1$ and $N = 2$ species of fermions on the square lattice with $\pi-$flux. The power-law exponent is considerably enhanced compared with the unprojected wave function (dashed line), consistent with predictions of spin liquid theory. 
     }
     \label{Piflux}
 \end{figure}

\section{Dirac Spin liquid}

A spin liquid state typically involves an emergent dynamical gauge field~\cite{motrunich2005,ran2007,hermele2008,Grover2010,xuVMC,becca,becca2,fisherboseliquid}, and the exact gauge group depends on the projection operator, as well as the ``mean field" state before the projection. 
When the emergent gauge symmetry is U(1), and the fermions $f_{i,\alpha}$ have Dirac nodes at the Fermi level, the low energy action describing the spin liquid is a QED$_3$ with Dirac fermion matter fields coupled to a $\U(1)$ gauge field~\footnote{The Gutzwiller wave function may not faithfully correspond to the exact full gauge theory. For example, Ref.~\cite{lesik} discussed a case where the Gutzwiller wave function does not capture the full fluctuations of gauge field. We will include a more detailed discussion between these two approaches in a future work. } 

QED$_3$ Dirac spin liquids have attracted great interests~\cite{hasting2000,hermele2005,ran2007,hermele2008,becca,becca2,hesl,hesl2,lauchlisl}. The dynamical gauge field $A_\mu$ may lead to confinement and hence destabilize the spin liquid states. In our construction, to ensure a deconfined phase of the gauge field, one can start with a state $|\Psi_0\rangle$ that is the ground state of $N$ flavors of degenerate fermions $c_{i,I}$ at half-filling, where $I = 1 \cdots N$. For example, one can start with a state of the alkaline earth cold atoms, which enjoy a large flavor symmetry~\cite{xusun}. Then eventually the effective Gutzwiller projection imposed on the mixed state in the doubled space is \beqn \sum_{I = 1 \cdots N} \hn_{i,I,\ua} + \hn_{i,I,\da} = N, \eeqn which ensures that $|\rho\rrangle$ is a wave function of an effective $\SU(2N)$ spin system, with self-conjugate representation of $\SU(2N)$ on each site.

Assuming the dispersion of $|\Psi_0\rangle$ has Dirac nodes in the momentum space with $N-$fold flavor degeneracy, the spin liquid that $|\rho\rrangle$ simulates is described by the following theory:
\begin{equation}
\cS = \int d^2x d\tau \sum_{v = 1}^{N_v} \sum_{a = 1}^{2N} \bar{\psi}_{a,v} \gamma_\mu (\partial_\mu - \ii A_\mu) \psi_{a,v} + \frac{1}{4e^2}F_{\mu\nu}^2. \label{qed1}
\end{equation}
Here $v$ labels the $N_v$ Dirac nodes/valleys in the momentum space. For example, if $|\Psi_0\rangle$ is the ground state of a nearest neighbor tight binding model on the honeycomb lattice at half-filling, or on the square lattice with $\pi-$flux per square, then $N_v = 2$, i.e. there are two independent Dirac cones in the momentum space. It is known that for large enough $N$, Eq.~\ref{qed1} describes a stable algebraic liquid state which is also a $(2+1)d$ conformal field theory. We refer to this liquid state in the doubled Hilbert space as the algebraic Choi-spin liquid.

Predictions for the Choi-spin liquid described above can be made based on our understanding of the spin liquid. Let us still start with a SU($2N$) spin liquid on the square lattice with $\pi-$flux, whose low energy physics is described by QED$_3$ in Eq.~\ref{qed1}. The local spin operator $\hat{S}^z_i = \sum_{I = 1 \cdots N} \hn_{i,I,\ua} - \hn_{i,I,\da}$ has a nonzero overlap with the fermion-bilinear composite field $\bar{\psi} S^z \mu^z \psi$, where $\mu^z = \pm 1$ denotes the two Dirac valleys. $\bar{\psi} S^z \mu^z \psi$ is an $\SU(2N)$ N\'{e}el order parameter. Then at long distance the spin correlation functions is given by \beqn \langle \hat{S}^z_0 \ \hat{S}^z_{\vect{x}} \rangle \sim (-1)^{\vect{x}} \frac{1}{|\vect{x}|^{2\Delta}} + \cdots. \label{corre} \eeqn Note that the sign of the correlation function above oscillates depending on the sublattice of $\vect{x}$. With sufficiently large $N$, the scaling dimension $\Delta$ can be computed using the standard $1/N$ expansion, and it is smaller than the scaling dimension of the free Dirac fermion~\cite{rantner}: \beqn \Delta = 2 - \frac{32}{3 N N_v
\pi^2} + O\left(\frac{1}{N^2}\right). \label{scaling} \eeqn There is also a non-oscillating (ferromagnetic) part of the correlation in Eq.~\ref{corre}, but it decays with a larger power $\Delta' = 2$, which is the same as that of the free Dirac fermion, as the ferromagnetic component of the spin density is a conserved quantity.

It was shown that~\cite{parapiflux}  a Gutzwiller-projected $1d$ free fermion wave function with a $\SU(N)$ symmetry generates accurate correlations predicted by $(1+1)d$ conformal field theories. And a Gutzwiller-projected $\pi-$flux state on the square lattice leads to enhanced power-law correlations, qualitatively consistent with the prediction of QED$_3$. We have conducted numerical study on the Gutzwiller projected $\pi-$flux state on the square lattice, indeed we see a power-law scaling with a considerably smaller scaling dimension compared with free Dirac fermion (Fig.~\ref{Piflux}), which is again qualitatively consistent with field theory predictions Eq.~\ref{scaling}.

The $\SU(2N)$ spin correlation function above can be used to make predictions for the following ``Renyi-2" correlation in the current context:
\beqn && \llangle \rho| \delta \hn_{0,\ua} \ \delta \hn_{\vect{x},\ua} |\rho\rrangle \sim \llangle \rho| \hat{S}^z_0 \ \hat{S}^z_\vect{x} |\rho\rrangle \cr\cr &\sim& \tr \left( \rho^2 \delta \hn_0 \ \delta \hn_{\vect{x}} \right) \sim (-1)^{\vect{x}} \frac{1}{|\vect{x}|^{2\Delta}}.\label{eq:su(2N)_renyi-2} \eeqn
Here $\delta \hn(\vect{x}) = \hn(\vect{x}) - N/2$, and we have used the fact that $\delta n_\ua(\vect{x}) \sim \hat{S}^z_{\vect{x}}/2$, since $\hn_{\ua}(\vect{x}) + \hn_{\da}(\vect{x})$ is a constant.
Since the Choi state $|\rho\rrangle $ has a $\SU(2N)$ symmetry, all the $\SU(2N)$ spin correlation function should have the same scaling as Eq.~\ref{eq:su(2N)_renyi-2}. For example, \beqn && \tr \left( \rho^2 \ (\hO_0)^I_J \ (\hO_{\vect{x}})^J_I \right) \label{renyi3} \eeqn is expected to decay in the same manner as Eq.~\ref{eq:su(2N)_renyi-2} for sufficiently large $N$, where $\hO^I_J = c^\dagger_{I} c_{J}$ with $I \neq J$ is a $\SU(N)$ operator that operates on the index $I = 1 \cdots N$. Note that $\hO$ does not change the total number configuration $\mathbf{n}$. In the SM we describe the experimental protocol to probe the correlation function predicted above.

\section{SU(2) gauge symmetry and superconductivity}

It is well-known in the field of spin liquid that (see for example Ref.~\cite{wensl}), a U(1) projection may actually lead to SU(2) gauge symmetry. In our context, for a class of free fermion states $|\Psi_0\rangle$ with $N = 1$ whose parent Hamiltonian $H_0$ only has real hopping $t_{ij}$, the seemingly $\U(1)$ gauge projection in the doubled space $\prod_i \hat{\calP}(\hn_{i,\ua} + \hn_{i,\da} = 1)$ would also lead to a SU(2) gauge symmetry. For example, the $\pi-$flux state with $N = 1$ in Fig.~\ref{Piflux} actually has SU(2) gauge symmetry. 

To expose the explicit $\SU(2)$ gauge symmetry, we no longer need the PH transformation of $H^*_0(c_{i,R})$. Instead, we just need to (trivially) relabel $L \ra 1$, $R \ra 2$, then the $\U(1)$ gauge projection $\hn_{i,1} - \hn_{i,2} = 0$ automatically implies a SU(2) gauge constraint: \beqn c^\dagger_{i,\alpha} \tau^l_{\alpha,\beta} c_{i,\beta} = 0, \ \ \ l = 1,2,3. \eeqn The reason is that, in the fock space of $c_{i,1}$ and $c_{i,2}$, the only states that survive the projection are $|0,0\rangle_i $, and $c_{i,1}^\dagger c_{i,2}^\dagger |0,0\rangle_i$, both are singlets under a local SU(2) rotation on $[c_{i,1}; c_{i,2}]$. 

There are a few important points to keep in mind:

(1) This SU(2) gauge structure {\it only} emerges when $N = 1$, i.e. when the original state $|\Psi_0\rangle$ is a spinless fermion state, and there are two flavors of fermions in the doubled Hilbert space. 

(2) The SU(2) gauge symmetry is {\it independent} from the effective SU(2) global spin symmetry, i.e. one can explicitly break the SU(2) spin symmetry, but still keep the SU(2) gauge symmetry.  

(3) The gauge symmetry may be broken down to a smaller gauge group, depending on the original state $|\Psi_0\rangle$, or its parent Hamiltonian $H_0$~\cite{wensl}. More specifically, the parent Hamiltonians for $|\rho_0\rrangle$ is $H_0(c_{i,L}) + H_0^\ast (c_{i,R})$. If $H_0$ is real (i.e. the hopping amplitudes $t_{ij}$ are real), then the SU(2) gauge symmetry is preserved in the infrared.

Motivated by experiments in candidate spin liquid materials, one of the spin liquid states discussed most is the ``spinon Fermi surface state"~\cite{kappa1,dmit1,dmit2,herb,zhousl,xusl1,xusl2,chensl,amperean,lee2005,motrunich2005}. A ``spinon Fermi surface" state can be naturally produced in our context when the original state $|\Psi_0\rangle$ is a spinless fermion state with a Fermi surface. As we mentioned above, when its parent Hamiltonian $H_0$ is real, the spinon Fermi surface state in the doubled Hilbert space is coupled to a $\SU(2)$ gauge field. Note that in our current case, the Fermi surface state after projection would most naturally have a SU(2) gauge symmetry as long as $H_0$ is real, rather than a $\U(1)$ gauge symmetry as was often discussed in the literature of spin liquids. 

A $(2+1)d$ Fermi surface coupled to a dynamical bosonic field (e.g. a dynamical gauge field) is a challenging theoretical problem in general, and it has continuously attracted enormous theoretical interests and efforts~\cite{hertz,millis,polchinski,nayak1,nayak2,mrossnfl,nflpair1,nflpair2,nflpair3,nflinstable,sslee,maxnematic}.In our current case, intuitively, the SU(2) gauge field would yield an attractive interaction between $c_1$ and $c_2$, and this attractive interaction may lead to pairing instability, i.e. in the doubled Hilbert space there could be condensate of the Cooper pair operator $\hat{\Delta} = c_1c_2$. A Cooper pair condensate leads to the following long-range correlation:
\begin{align}
\llangle \rho_\infty| (c_1 c_2)_0 \ (c_1 c_2)^\dagger_{\vect{x}} |\rho_\infty\rrangle &= \tr\left( c^\dagger_0 \rho_\infty c_0 \ c^\dagger_{\vect{x}} \rho_\infty c_{\vect{x}} \right) \nonumber\\
&\xrightarrow{\vect{x}\to \infty} \mathrm{Const}. \label{SC}
\end{align}
Superconductivity of matter fields coupled to a non-Abelian gauge field is referred to as color superconductivity (for reviews, see Ref.~\cite{colorSC1,colorSC2}). The superconductivity can be obtained through an $\epsilon$-expansion in theory~\cite{SU2parton8,mandal}, and eventually an extrapolation to $\epsilon = 1$. An experimental realization of the state in the quantum simulator can serve as a test for this theoretical prediction. 

We have also numerically studied the projected Fermi surface state with two flavors of fermions, and indeed we observe a long range correlation of 
Cooper pairs (Fig.~\ref{SC}), consistent with our theoretical prediction. 
\begin{figure}[h!]
     \includegraphics[width=0.8\linewidth]{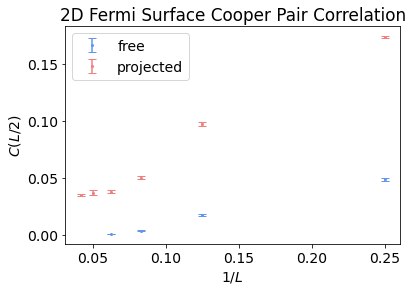}
     \caption{The projected free fermion wave function with two-flavors of fermions and a Fermi surface. There is a long range correlation of Cooper pair, as expected. The numerics is performed on a square lattice tight binding model with nearest neighbor and 2nd neighbor hopping, $t_2/t_1 = 0.2$. We fix the system size $L \times L$, and measure Cooper pair correlation at $r = L/2$. }
     \label{SC}
 \end{figure}

In the doubled space, a Cooper pair condensate spontaneously breaks the charge $\U(1)_e$ symmetry, which is also the so-called ``strong" $\U(1)$ symmetry of the density matrix~\cite{buvca2012note,groot}. The operator $c^\dagger_{\vect{x}} \rho c_{\vect{x}}$ transforms nontrivially under the strong $\U(1)$ symmetry, but invariant under the ``weak" $\U(1)$ symmetry. Hence the long-range correlation of Eq.~\ref{SC} implies a ``strong-to-weak" spontaneous breaking of the $\U(1)$ symmetry in the mixed state. 


\section{Summary and Discussion}

In this work, we demonstrated that starting with a simple fermion wave function, the steady state density matrix of the Lindbladian evolution, or the ensemble generated from strong measurement of local densities, is a Gutzwiller projected wave function in the doubled Fock space. The Gutzwiller projection has been broadly used as a numerical trial wave function of spin liquid. We propose that this new construction of spin liquids can be realized in real experimental platforms, e.g. the Fermi gas microscope. 
We also predict that in certain scenarios, the constructed mixed state has a strong-to-weak spontaneous $\U(1)$ symmetry breaking, which corresponds to a superconductor in the doubled Fock space. Predictions made by our understanding of spin liquids, as well as the detailed measurement protocol of our predictions are described in the SM. 

So far we have assumed that $|\Psi_0\rangle$ is the ground state of a free fermion tight-binding model. We expect most of our predictions to be robust against a certain amount of short-range interaction in $H_0$. For example, when $H_0$ is a nearest neighbor tight-binding model of $N$ flavors of fermions $c_{I}$ on the honeycomb lattice at half-filling, the Choi states $|\rho_0\rrangle $ and $|\rho\rrangle $ should have an exact $\SU(2N)$ symmetry, which guarantees that the correlations in Eq.~\ref{eq:su(2N)_renyi-2} and Eq.~\ref{renyi3} take the same form. If $H_0$ has weak short-range interactions, then the Choi states no longer have an exact $\SU(2N)$ symmetry. We expect the Dirac fermion $\psi$ in Eq.~\ref{qed1} to inherit extra interactions from $H_0$. But since short-range interactions in Eq.~\ref{qed1} are perturbatively irrelevant for large enough $N$, the $\SU(2N)$ symmetry is expected to emerge in the infrared, and the correlations in Eq.~\ref{eq:su(2N)_renyi-2} and Eq.~\ref{renyi3} should still have the same scaling dimension $\Delta$ for large $\vect{x}$. For smaller $N$, some short-range interactions can become relevant due to the gauge fluctuation~\cite{Kaveh_2005,xu4f1,xu4f2,pufu4f,he4f,Braun_2014}.
For a non-interacting initial state $\rho_0 = \ketbra{\Psi_0}$, the correlator in the spin liquid state $\kket{\rho}$ actually stems from observables in the free fermion wave function $\kket{\rho_0}$, which can be computed on classical computers. For an initial gapless state with short-range interactions, we propose a combined classical-classical correlator and quantum-classical correlator to make predictions on Eq.~\ref{eq:su(2N)_renyi-2}, Eq.~\ref{renyi3} for smaller $N$ (please refer to the SM). 

In this work we have focused on mixed states prepared from initial fermion states $|\Psi_0\rangle$ that are {\it gapless}. Gapped fermion states can also be very interesting -- even free fermion insulators can have nontrivial topological nature, e.g. the Chern insulator. The Choi-representation of the Chern insulator in the doubled space is a quantum spin Hall state, and the procedure proposed in this work will lead to a Gutzwiller projection of the QSH wave function. We leave the generalizations of our work to topological insulators, including a discussion of the relation between Gutzwiller wave function and gauge theory to a future work. 

The authors thank Leon Balents, Chao-Ming Jian, David Weld, \yimu{and Muqing Xu} for very helpful discussions. C.X. is supported by the Simons Foundation International through the Simons Investigator grant. Y.B. and T.K. are supported in part by grant NSF PHY-2309135 to the Kavli Institute for Theoretical Physics (KITP).
Y.B. is supported in part by the Gordon and Betty Moore Foundation Grant No. GBMF7392 to the KITP.
T.K. is supported in part by the Gordon and Betty Moore Foundation Grant No. GBMF8690 to the KITP.


\appendix

\section{Measurement of R\'enyi-2 correlation}

In this appendix, we develop protocols to measure the R\'enyi-2 correlations in the Gutzwiller projected doubled state $\kket{\rho}$, 
\beqn \llangle \rho| \delta \hn_{0,\ua} \ \delta \hn_{\vect{x},\ua} |\rho\rrangle &\sim& \llangle \rho| \hat{S}^z_0 \ \hat{S}^z_\vect{x} |\rho\rrangle \cr\cr &\sim& \tr \left( \rho^2 \ \delta \hn_0 \ \delta \hn_{\vect{x}} \right), \label{renyi2} \eeqn here $\delta \hn_{\vect{x}} = \hn_\vect{x} - \bar{n}$. Our goal is to verify that $\kket{\rho}$ is an algebraic spin liquid state, meaning that the correlator above decays as $\sim 1/|\vect{x}|^{2\Delta}$, and we would like to extract the scaling dimension $\Delta$. 

If state $\kket{\rho}$ is prepared from a {\it free} fermion state $|\Psi_0\rangle$ after measuring the fermion density configurations, i.e. $\rho$ is a free fermion state subject to complete dephasing, the R\'enyi-2 correlator can be computed on a classical computer.
However, if $|\Psi_0\rangle$ is an interacting fermion state,
directly computing the R\'enyi-2 correlator is generally challenging.
%
We here provide a scheme to verify the spin liquid prediction prepared from fermion states {\it with or without } short-range interactions. We note that experimentally one can prepare two identical copies of the state and measure the SWAP operator to evaluate R\'enyi-2 quantities~\cite{daley2012measuring,pichler2013thermal,kaufman2016quantum}. We would like to propose a different method that avoids duplicating the system, and it only requires making measurement of local density configurations. 

Our protocol makes use of the decomposition of the density matrix $\rho $ as an ensemble of pure states generated by the measurements of local fermion occupation $\hn_i = \sum_\alpha \hn_{i,\alpha}$
\begin{align}
\rho &= \sum_{\mathbf{n}} q_{\mathbf{n}} \ketbra{\Psi_{\mathbf{n}}} = \sum_{\mathbf{n}} \hat{P}_{\mathbf{n}} \ketbra{\Psi_0} \hat{P}_{\mathbf{n}},
\end{align}
where $\mathbf{n}$ is a vector that labels the fermion occupation on each site, $\ket{\Psi_{\mathbf{n}}}$ is the normalized post-measurement state, $\hat{P}_{\mathbf{n}}$ is the projector onto the subspace with occupation number $\mathbf{n}$, and $q_{\mathbf{n}} = \bra{\Psi_0}\hat{P}_{\mathbf{n}}\ket{\Psi_0}$ is the probability of measurement outcome $\mathbf{n}$. To proceed, we need to first define the ``classical-classical" correlator, and the ``quantum-classical" correlator, motivated from Ref.~\cite{altman1}. 


\subsection{The ``classical-classical" correlator}

We first consider a ``classical-classical" (CC) correlator defined with the projected {\it free fermion} state $\kket{\tilde{\rho}}$
\begin{align}
\llangle \delta \hn_0 \ \delta \hn_{\vect{x}} \rrangle_{\tilde{\rho}} = \frac{\tr \tilde{\rho}^2 \ \delta \hn_0 \ \delta \hn_{\vect{x}}}{\tr\tilde{\rho}^2} = \frac{\sum_{\mathbf{n}} p_{\mathbf{n}}^2 \ \delta n_0 \ \delta n_{\vect{x}}}{\sum_{\mathbf{n}} p_{\mathbf{n}}^2}, \label{CC}
\end{align}
where $\llangle \delta \hn_0 \ \delta \hn_{\vect{x}} \rrangle_{\tilde{\rho}} := \bbra{\tilde{\rho}} \delta \hn_0 \ \delta \hn_{\vect{x}}\kket{\tilde{\rho}}$, $p_{\mathbf{n}} = \langle \tilde{\Psi}_0 |\hP_{\mathbf{n}}|\tilde{\Psi}_0\rangle $ denotes the probability of outcome $\mathbf{n}$ in the {\it free fermion} state $|\tilde{\Psi}_0\rangle$.
%
\cx{We stress that $|\Psi_0\rangle$ in the previous subsection is the ground state of a gapless Hamiltonian $H_0$ (which can include short-range interactions) for $N-$flavors of fermions $c_{I}$ with an $\SU(N)$ symmetry, while $|\tilde{\Psi}_0\rangle$ is the free fermion ground state of the non-interacting part (fermion-bilinear part) of $H_0$.}
Specifically, we first sample from $|\tilde{\Psi}_0\rangle$ in the local occupation basis independently for $M$ times, which can be efficiently performed on a classical computer~\cite{valiant2001quantum,terhal2002classical,knill2001fermionic}.
Given the outcome $\mathbf{n}_m$ obtained in the $m$-th sample, we compute $p_{\mathbf{n}_m}$.
Taking $M$ independent samples, one can obtain an estimate of the CC-correlator, i.e.
\begin{align}
\llangle \delta \hn_0 \ \delta \hn_{\vect{x}} \rrangle_{\tilde{\rho}} = \frac{\sum_{m = 1}^M p_{\mathbf{n}_m} \delta n_{0,m} \ \delta n_{\vect{x},m}}{\sum_{m = 1}^M p_{\mathbf{n}_m}}.\label{cc_corr}
\end{align}
Using the same set of samples, we can compute $\langle \tilde{\Psi}_0| \hP_{\mathbf{n}_m} (\hO_0)^I_J \ (\hO_{\vect{x}})^J_I |\tilde{\Psi}_0\rangle$ on the classical computer and obtain the R\'enyi-2 correlator in Eq.14,
\beqn
&& \llangle (\hO_0)^I_J (\hO_{\vect{x}})^J_I \rrangle_{\tilde{\rho}} \cr\cr &=& \frac{\sum_{m = 1}^M \langle \tilde{\Psi}_0| \hP_{\mathbf{n}_m} (\hO_0)^I_J(\hO_{\vect{x}})^J_I |\tilde{\Psi}_0\rangle }{\sum_{m = 1}^M p_{\mathbf{n}_m}}. \label{cc_corr2}
\eeqn
Here, we use the fact that $(\hO_{\vect{x}})_J^I = c_{I,\vect{x}}^\dagger c_{J,\vect{x}}$ does not change the fermion occupation $\mathbf{n}$. 
%
In Eq.~\ref{cc_corr} and Eq.~\ref{cc_corr2}, since both sampling and computation are performed on classical computers, we refer to them as the classical-classical (CC) correlator.

\subsection{The ``quantum-classical" correlator}


As the second step, we consider the following ``quantum-classical" (QC) correlator, which is defined as the R\'enyi-2 correlator in the doubled state associated with \beqn \sigma = \sum_{\mathbf{n}} \hP_{\mathbf{n}} \ket{\Psi_0}\langle\tilde{\Psi}_0| \hP_{\mathbf{n}}, \label{sigma} \eeqn
\begin{align}
\kket{\sigma} = \prod_i \hat{\mathcal{P}}(n_{i,L}=n_{i,R}) |\Psi_0\rangle_L |\tilde{\Psi}\rangle_R.
\end{align} \cx{Note that in the doubled space, $|\Psi_0\rangle_L$ is the {\it experimentally prepared} gapless fermion state which can be interacting; while $|\tilde{\Psi}\rangle_R$ is the {\it free fermion state} that is the ground state of the noninteracting part of the parent Hamiltonian of $|\Psi_0\rangle_L$. $|\sigma\rrangle$ is a Gutzwiller projection on $2N$ flavors of fermions, but $N$ flavors of the fermions (in $|\tilde{\Psi}_0\rangle$) are non-interacting. }

\yimu{The R\'enyi-2 correlator} of $\sigma$ then takes the form
\begin{align}
\llangle \delta \hn_0 \ \delta \hn_{\vect{x}} \rrangle_\sigma = \frac{\tr \sigma^\dagger \sigma \ \delta \hn_0 \ \delta \hn_{\vect{x}}}{\tr\sigma^\dagger \sigma} = \frac{\sum_{\mathbf{n}} q_{\mathbf{n}} p_{\mathbf{n}} \ \delta n_0 \ \delta n_{\vect{x}}}{\sum_{\mathbf{n}} q_{\mathbf{n}}p_{\mathbf{n}}}, \label{QC}
\end{align}
where $q_\mathbf{n}$ is the probability of fermion occupation $\mathbf{n}$ in the interacting wave function $\ket{\Psi_0}$ prepared experimentally in quantum simulator.
The quantity can be evaluated by first sampling from the interacting wave function $\ket{\Psi_0}$ on a quantum simulator (governed by the probability distribution $\{q_\mathbf{n}\}$) and then computing the corresponding $p_{\mathbf{n}}$ in the free fermion state $|\tilde{\Psi}_0\rangle$ on a classical computer.
\yimu{We then have the following estimate of the quantum-classical correlator}
\begin{align}
\llangle \delta \hn_0 \delta \hn_{\vect{x}} \rrangle_{\sigma} = \frac{\sum_{m = 1}^M p_{\mathbf{n}_m} \delta n_{0,m}\delta n_{\vect{x},m}}{\sum_{m = 1}^M p_{\mathbf{n}_m}}. \label{QCe}
\end{align}

\subsection{Comparison between the CC and QC correlators}

To make predictions regarding the desired quantity Eq.~\ref{renyi2}, we will need both the CC-correlator Eq.~\ref{CC} and QC correlator Eq.~\ref{QC}. Here we argue that, if the QC and CC correlators exhibit the same power-law scaling, this strongly indicates that the experimentally generated $\kket{\rho}$ is also a spin liquid state, with the same power-law correlations.
%


It is well-known that, for large enough $N$, all the perturbations of short-range four-fermion interactions in the QED$_3$ theory (Eq.11 of the main text) are irrelevant. But with smaller $N$, some short-range interactions can become relevant and destabilize the spin liquid described by Eq.11. The scaling dimensions of the short-range interactions depend on $N$, and also the symmetry of the four fermion interactions~\cite{Kaveh_2005,xu4f1,xu4f2,pufu4f,he4f,Braun_2014}.

Hence we need to compare the exact symmetry of the CC-correlator generated by $|\tilde{\rho} \rrangle$, QC-correlator generated by $|\sigma \rrangle$, and the quantum-quantum (QQ) correlator Eq.~\ref{renyi2} which corresponds to the experimentally prepared $|\rho\rrangle$. As an example, let us still start with a $H_0$ with $N$ flavors of fermions and nearest neighbor hopping on the honeycomb lattice, as well as some local density-density interactions preserving the $\SU(N)$ flavor symmetry. The CC-correlator based on the projected non-interacting state $|\tilde{\rho}\rrangle$ has an exact $\SU(2N)$ symmetry on the lattice, as we discussed in the main text; the QQ-correlator which corresponds to the projected interacting state $|\rho\rrangle$ has an exact $(\SU(N)_L \times \SU(N)_R) \rtimes Z_2$ symmetry, where the $Z_2$ exchanges the left and right space. While the QC-correlator evaluated with the Choi state $|\sigma\rrangle$ has the {\it lowest} $\SU(N)_L \times \SU(N)_R$ symmetry. Though all three kinds of correlators may correspond to field theory Eq.11 with perturbations (extra four-fermion interactions), since $|\sigma\rrangle$ has the lowest symmetry, the field theory corresponding to $|\sigma\rrangle$ has {\it the most} four-fermion terms in addition to Eq.11. If the QC and the CC-correlator yield the same power-law decay with the same scaling dimension, the only natural possibility is that the local interactions in both QC and the CC-correlators are irrelevant; this would imply that the local interactions in QQ-correlator are also irrelevant since the QQ-correlator has higher symmetry than the QC-correlator, and must have fewer perturbations of local interaction. This means that $|\rho \rrangle$ prepared experimentally should indeed be an algebraic spin liquid state controlled by the fixed point described by Eq.11. 

\cx{We can also evaluate the Cooper pair correlation in Eq.~17 in a similar way. For example, the CC-correlator of Eq.17 reads}
\begin{align}
\frac{\tr c_0^\dagger \tilde{\rho} c_0 c_{\vect{x}}^\dagger \tilde{\rho} c_{\vect{x}}}{\tr \tilde{\rho}^2} = \frac{\sum_{m = 1}^M p_{\mathbf{n}'_{m}}}{\sum_{m = 1}^M p_{\mathbf{n}_m}},
\end{align}
where, for an outcome $\mathbf{n}_m = (n_{0,m}, \cdots, n_{\vect{x},m}, \cdots)$, $\mathbf{n}'_m = (n_{0,m}+1,\cdots,n_{\vect{x},m}-1,\cdots)$.
In this case, we note that the fermion on each site has a single flavor ($N = 1$), and the mixed state $\tilde{\rho}$ is classical, i.e. a diagonal matrix. \cx{The QC-correlator of Eq.17 can be constructed accordingly.} The only difference is that, for an experimentally generated configuration $\mathbf{n}_m$, we need to compute the probability $p_{\mathbf{n}'_m}$ from a classical computer, for a slightly different configuration $\mathbf{n}'_m$. 

\subsection{Summary of Experimental Protocol}

Here we make a summary of the proposed experimental protocol for measuring the desired quantity Eq.~\ref{renyi2}:

\begin{itemize}
    
    \item Prepare a fermion state $|\Psi_0\rangle$ (cold atoms in optical lattice) that is the ground state of Hamiltonian $H_0$. $H_0$ can have short-range interactions, and its spectrum is gapless.   
    
    \item In each experimental run, measure the fermion density on each site, obtain configuration $\mathbf{n}_m$.
    
    \item Input $\mathbf{n}_m$ in computer, find $p_{\mathbf{n}_m} = \langle \tilde{\Psi}_0| \hat{P}_{\mathbf{n}_m} | \tilde{\Psi}_0 \rangle$, where $|\tilde{\Psi}_0\rangle$ is the free fermion ground state of the noninteracting part of $H_0$.                         
    \item We average $p_{\mathbf{n}_m} \delta n_{0,m}\delta n_{\vect{x},m}$ over all experimental runs like Eq.~\ref{QCe}, and obtain the QC-correlator.

    \item We separately compute the CC-correlator using $| \tilde{\Psi}_0 \rangle$. If the CC-correlator and QC-correlator have the same power-law scaling, based on the arguments in the previous subsection, we verify that the experimentally prepared $\kket{\rho}$ is a spin liquid state, and its scaling is given by the QC and CC correlator.

\end{itemize}


\section{Other gapless liquid states}

\subsection{Spinon Fermi surface state}

The proposed experimental protocol discussed in the previous section also allows us to experimentally extract information of other gapless liquid states, for example, the spinon Fermi surface state, and the $d$-wave Bose liquid phase~\cite{fisherboseliquid,fisherboseliquid2}. 

The spinon Fermi surface state is among the most broadly studied spin liquid states, due to the compelling potential connection to condensed matter experiments. In our context, to access the physics of the gapless spinon Fermi surface state, we can follow the following protocol:

\begin{itemize}
    
    \item Prepare a spinless fermion state $|\Psi_0\rangle$ that is a gapless state at half-filling with a Fermi surface, for example, it can be chosen as the ground state of the tight-binding model on the triangular lattice at half-filling. $|\Psi_0\rangle$ can have short-range interactions.   
    
    \item Again, in each experimental run, measure the fermion density on each site, obtain configuration $\mathbf{n}_m$.

    \item For each experimentally produced configuration $\mathbf{n}_m$, we define $\tilde{\mathbf{n}}_m$, and $\tilde{n}_{\vect{x},m} = 1 - n_{\vect{x},m} $. 
    
    \item Input $\tilde{\mathbf{n}}_m$ into computer, find $p_{\tilde{\mathbf{n}}_m} = \langle \tilde{\Psi}_0| \hat{P}_{\tilde{\mathbf{n}}_m} | \tilde{\Psi}_0 \rangle$, where $|\tilde{\Psi}_0\rangle$ is the free fermion ground state of the noninteracting part of the parent Hamiltonian of $|\Psi_0\rangle$.  
    
    \item We average $p_{\tilde{\mathbf{n}}_m} \delta n_{0,m}\delta n_{\vect{x},m}$ over all experimental runs, and this quantity should converge to the spin correlation function of the following state: \beqn \frac{\sum_m p_{\tilde{\mathbf{n}}_m} \delta n_{0,m}\delta n_{\vect{x},m}}{\sum_m p_{\tilde{\mathbf{n}}_m} } \sim \llangle \sigma | S^z_0 \ S^z_{\vect{x}} | \sigma \rrangle, \eeqn 
    and state $| \sigma \rrangle$ is \beqn |\sigma \rrangle \sim \prod_{i} \hat{\calP}(n_{i,L} + \tilde{n}_{i,R} = 1 ) |\Psi_0\rangle_L |\tilde{\Psi}_0\rangle_R. \eeqn 

\end{itemize}

We can view $|\Psi_0\rangle$ and $|\tilde{\Psi}_0\rangle$ as the wave function for spin-up and spin-down spinons. $|\Psi_0\rangle$ and $|\tilde{\Psi}_0\rangle$ have the same Fermi surface. If $|\Psi_0\rangle$ is a free fermion state with a finite Fermi surface, e.g. the ground state of a half-filling tight-binding model on the triangular lattice, the constructed $|\sigma \rangle$ has a spin SU(2) symmetry. If $|\Psi_0\rangle$ has short-range interactions, $|\sigma \rrangle$ is a special state where the spin-up and down spinons have different interactions, but still the same Fermi surface. 

\subsection{$d$-wave Bose Liquid state}

Ref.~\cite{fisherboseliquid,fisherboseliquid2} discussed an exotic $d$-wave Bose liquid state for interacting bosons. The construction started with a parton formalism \beqn b_i = d_{i,x} d_{i,y}, \eeqn where $d_{i,x}$ and $d_{i,y}$ are fermionic partons with strong hoppings along $x$ and $y$ directions respectively. The $d$-wave Bose liquid state corresponds to the mean field state where the two partons each has their own Fermi surfaces with anisotropic shapes. 

Now let us assume that a state $|\Psi_{0,x}\rangle$ is prepared in experiment, which is a fermion state with stronger hopping along the $x$ direction, and weaker hopping along the $y$ direction. Again, in each experimental run, one measures the fermion density $|\Psi_{0,x}\rangle$ on each site, and obtain configuration $\mathbf{n}_m$. We then input $\mathbf{n}_m$ in another state $|\tilde{\Psi}_{0,y}\rangle$ in computer, and find $p_{\mathbf{n}_m} = \langle \tilde{\Psi}_{0,y}| \hat{P}_{\mathbf{n}_m} | \tilde{\Psi}_{0,y} \rangle$. Here $|\tilde{\Psi}_{0,y}\rangle$ is a state with stronger hopping along $y$, and weaker hopping along $x$. We then average quantity $p_{\mathbf{n}_m} \delta n_{0,m}\delta n_{\vect{x},m}$ over experimental runs, this quantity should converge to the density correlation of the following state \beqn 
|\sigma \rrangle \sim \hat{\calP}(\hat{n}_{i,x, L} = \hat{n}_{i,y, R}) |\Psi_{0,x}\rangle_L |\tilde{\Psi}_{0,y}\rangle_R. \eeqn $|\sigma \rrangle$ is precisely the Gutzwiller construction of the $d$-wave Bose liquid state.

\section{More general QCD$_3$ Choi-spin liquids}

In the main text We have discussed that the QCD with an emergent SU(2) gauge field may be realized in the Choi state after a Lindbladian evolution when the jump operators are local density operators. In this section, we will demonstrate that QCD with a more general $\SU(k)$ gauge field can also be realized in a similar manner, by choosing a different set of jump operators. Now let us consider the initial state $|\Psi_0 \rangle$ which is a free fermion state of $kN$ flavors of fermions. We label the fermions as $c_{i,l,J}$, where $l = 1 \cdots k$ and $J = 1 \cdots N$. We choose the jump operator as \beqn L^\mu_i = \hat{T}^\mu_i = c^\dagger_{i,l,J} T^\mu_{ll'} c_{i,l',J}, \eeqn where $T^\mu$ with $\mu = 1 \cdots k^2 - 1$ are the generators of $\SU(k)$. Please note that $\hat{T}^\mu_i$ are all Hermitian operators. We can design the Lindbladian in the Choi-Jamio\l kowski representation as 
\begin{equation}
\cL = \sum_{i,\mu} - \frac{1}{2}(\hat{T}^\mu_{i,1} - (\hat{T}^\mu_{i,2})^T)^2 = \sum_{i,\mu} - \frac{1}{2}(\hat{T}^\mu_{i,1} - (\hat{T}^\mu_{i,2})^\ast)^2.
\end{equation}
Here $ (\hat{T}^\mu_{i,2})^\ast = c^\dagger_{i,l,J} (T^\mu_{ll'})^\ast c_{i,l',J} $. 

The fermions in the left and right spaces form a pair of complex-conjugate representations of $\SU(k)$, and in the long-time limit, the Lindbladian acts as a $\SU(k)$ Gutzwiller projection in the doubled Hilbert space, as it demands $\hat{T}^\mu_{i,1} - (\hat{T}^\mu_{i,2})^\ast = 0$ on every site: \beqn |\rho \rrangle = \prod_i \hat{\calP}(\SU(k))_i |\rho_0\rrangle, \eeqn where $\hat{\calP}(\SU(k))_i$ is a projection to the $\SU(k)$ singlet on every site $i$. It is hence expected that the steady state is a QCD with $2N$ flavors of fermions coupled with a $\SU(k)$ gauge field. Note that in $|\rho \rrangle$ $N$ flavors of the fermions (each carrying a ``color" index $l = 1 \cdots k$) form the fundamental rep of the $\SU(k)$, while the other $N$ flavors form an anti-fundamental representation.


\section{Doubled space representation of fermion density matrix}\label{app:fermionic_double_space}

Here, we detail the mapping from the fermion density matrix to the pure state in the doubled Fock space~\cite{bao2021symmetry}.
The mapping is slightly more complicated than mapping the density operator in quantum spin models to the doubled space because two copies of the fermionic operators in the doubled Fock space have mutual fermionic statistics. 

We first express the fermionic density matrix in the Fock basis as
\begin{align}
\rho(c_i) &= \sum_{\mathbf{n}_L, \mathbf{n}_{R}} \rho_{\mathbf{n}_L, \mathbf{n}_{R}} \ket{\mathbf{n}_L}\bra{\mathbf{n}_{R}} \nonumber \\
&= \sum_{\mathbf{n}_L, \mathbf{n}_{R}} \rho_{\mathbf{n}_L, \mathbf{n}_{R}} \calO^\dagger_{\mathbf{n}_L}\ketbra{\text{vac}} \calO_{\mathbf{n}_R},
\end{align}
where $\mathbf{n}_{L(R)} = (\{n_{i,L(R)}\})$ is a binary vector representing the occupation number of fermionic modes, $\calO^\dagger_{\mathbf{n}_L} := \prod_{i = 1}^N (c_i^\dagger)^{n_{i,L}}$, and $\calO_{\mathbf{n}_R} := \prod_{i = N}^1 (c_{i})^{n_{i,R}}$.

We now map the fermionic density matrix $\rho$ to a pure state $\kket{\rho}$ in the doubled Fock space which contains two copies of the fermionic mode $c_{i,L}$ and $c_{i,R}$.
We identify the fermonic mode $c_i$ in $\rho$ with $c_{i,L}$ in the doubled Fock space and obtain the doubled state $\kket{\rho}$ by applying $\rho(c_{i,L})$ as an operator to a reference state in the doubled Fock space, i.e. $\kket{\rho} := \rho(c_{i,L}) \kket{\mathcal{I}}$.
%
%
Here, we choose the following reference state
\begin{align}
\kket{\mathcal{I}} :=& \sum_{\mathbf{n}} \prod_{i=1}^N (c_{i,L}^\dagger)^{n_{i}} (c_{i,R}^\dagger)^{n_{i}} \kket{\text{vac}} \nonumber\\
=& \prod_{i=1}^N \left(1 + c_{i,L}^\dagger c_{i,R}^\dagger\right) \kket{\text{vac}}.
\end{align}
Accordingly, the doubled state $\kket{\rho}$ takes the form
\begin{align}
\kket{\rho} = \sum_{\mathbf{n}_L, \mathbf{n}_{R}} \rho_{\mathbf{n}_L, \mathbf{n}_{R}}\left(\calO^\dagger_{\mathbf{n}_L}\right)_L\left(\calO^T_{\mathbf{n}_R}\right)_R\kket{\text{vac}}.
\end{align}
Here, we introduce the transpose operation $(\cdot)^T$, which acts linearly on the fermionic operators,
\begin{gather}
c_i^T = c_i^\dagger, \quad (c_i^\dagger)^T = -c_i, \\
\left(\sum_{i} \alpha_i c_{i} + \beta_i c_{i}^\dagger\right)^T = \sum_i \alpha_i c_{i}^\dagger - \beta_i c_{i}.
\end{gather}
When acting on a product $\calO_{\mathbf{n}_1}\calO_{\mathbf{n}_2}$ with $\calO_{\mathbf{n}_{1,2}}$ individually being a product of fermionic operators defined above, we have 
\begin{align}
(\mathcal{O}_{\mathbf{n}_1}\mathcal{O}_{\mathbf{n}_2})^T = \sgn(\mathcal{O}_{\mathbf{n}_1},\mathcal{O}_{\mathbf{n}_2}) \mathcal{O}_{\mathbf{n}_2}^T \mathcal{O}_{\mathbf{n}_1}^T,
\end{align}
where $\sgn(\mathcal{O}_{\mathbf{n}_1}, \mathcal{O}_{\mathbf{n}_2}) = \pm 1$, it takes $-1$ value only if both $\mathcal{O}_{\mathbf{n}_1}$ and $\mathcal{O}_{\mathbf{n}_2}$ are fermionic (i.e. contain odd number of fermion operators).
One can verify that for any operator $(\calO^T)^T = \calO$.
According to this definition, we have the following relation for operators acting on the reference state
\begin{align}
    \mathcal{O}_L \kket{\mathcal{I}} = (\mathcal{O}^T)_R \kket{\mathcal{I}}.
\end{align}


In this work, we focused on an initial pure Gaussian (noninteracting) fermionic state evolved under the Lindblad equation.
Specifically, the initial pure state is $\rho_0 = \ketbra{\Psi_0}$ with
\begin{align}
\ket{\Psi_0} = \prod_{k = 1}^{N} (c_k^\dagger)^{n_k} \ket{\text{vac}}.
\end{align}
Here, $c_k^\dagger = \sum_i \alpha_i^* c_i^\dagger$ ($c_k = \sum_i \alpha_i c_i$) is a linear combination of fermionic creation operators on each site, e.g. the creation operator of fermion with momentum $k$.
The corresponding doubled state is given by
\begin{align}
    \kket{\rho_0} = \prod_{k=1}^N \left(c_{k,L}^\dagger c_{k,R}^T\right)^{n_k} \kket{\text{vac}}.
\end{align}
Here, $c_k^T = \sum_i \alpha_i c_i^\dagger$.
Accordingly, the Lindblad evolution in the doubled Fock space is given by
\begin{align}
\partial_t \kket{\rho} = \left(\sum_i (L_i)_L (L_i^\dagger)_R^T - \frac{1}{2} (L_i^\dagger L_i)_L -\frac{1}{2} (L_i^\dagger L_i)_R^T\right) \kket{\rho}.
\end{align}

Before closing, we remark that the choice of the reference state is not unique.
Alternatively, one can map $\rho \mapsto \rho(c_{i,L})\kket{\mathcal{I}'}$ with
\beqn
    \kket{\mathcal{I}'} :=& \sum_{\mathbf{n}} \prod_{i=1}^N (c_{i,L}^\dagger)^{n_{i}} \prod_{i=1}^N (c_{i,R}^\dagger)^{n_{i}} \kket{\text{vac}}. \eeqn 
%
However, such a choice is not convenient when considering the symmetry transformation that rotates between $c_{i,L}^\dagger$ and $c_{i,R}^\dagger$.
More specifically, the reference state $\kket{\mathcal{I}}$ has an $\SU(2)$ symmetry generated by $\hat{S}^\mu := \sum_i \frac{1}{2} \mathbf{c}^\dagger_i \sigma^\mu \mathbf{c}_i$, where $\mathbf{c}^\dagger_i = [c_{i,L}^\dagger, c_{i,R}^\dagger]$.
Such a symmetry would have to act nonlocally in the alternative reference state $\kket{\mathcal{I}'}$ due to the mutual fermionic statistics between $c_{i,L}$ and $c_{j,R}$. 
%
To make this $\SU(2)$ symmetry explicit and draw a direct connection to Gutzwiller projected wave function, we choose the reference state $\kket{\mathcal{I}}$ throughout this paper.


\bibliography{Lin}

\end{document}